\documentclass[journal=jpcbfk,manuscript=article]{achemso}
\setkeys{acs}{articletitle=true}
\usepackage{amsmath,amssymb}

\usepackage{graphicx}
\usepackage{dcolumn}
\usepackage{bm}
\usepackage{booktabs}
\usepackage{siunitx}
\usepackage{xcolor}
\usepackage{array}
\usepackage{hyperref}
\providecommand{\onlinecite}[1]{\citenum{#1}}
\newenvironment{ruledtabular}{\setlength{\tabcolsep}{3.5pt}}{}

\newcommand{\Tz}{\ensuremath{\mathrm{T}^0}}
\newcommand{\Tt}{\ensuremath{\mathrm{T}_2}}
\newcommand{\Tm}{\ensuremath{\mathrm{T}^-}}
\newcommand{\Fb}{\ensuremath{\mathrm{F}_\mathrm{b}}}
\newcommand{\muex}{\ensuremath{\mu_\mathrm{ex}}}

\title{Solubility and Diffusion of Hydrogen Isotopes and Helium in Molten FLiBe}

\author{Ayobami Daramola}
\email{ayobami.daramola@ukaea.uk}
\author{Vladimir Martis}
\affiliation{United Kingdom Atomic Energy Authority, Culham Campus, Abingdon, Oxfordshire OX14 3DB, United Kingdom}
\author{Graeme J. Ackland}
\affiliation{Centre for Science at Extreme Conditions and School of Physics and Astronomy, University of Edinburgh, Edinburgh, United Kingdom}
\author{Duc Nguyen-Manh}
\affiliation{United Kingdom Atomic Energy Authority, Culham Campus, Abingdon, Oxfordshire OX14 3DB, United Kingdom}
\alsoaffiliation{Department of Materials, University of Oxford, Oxford OX1 3PH, United Kingdom}

\begin{document}

\begin{tocentry}
\includegraphics[width=3.25in]{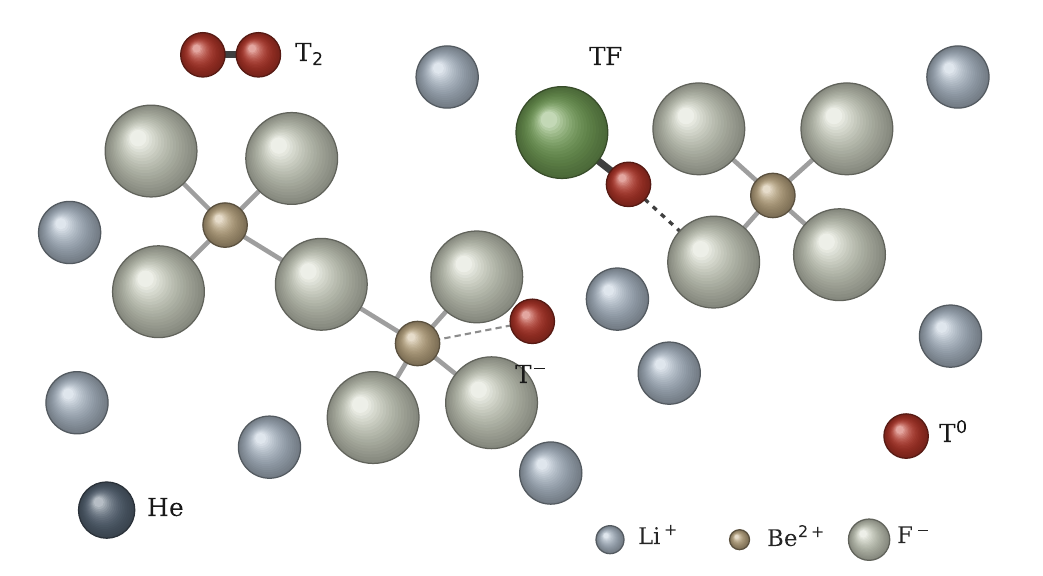}
\end{tocentry}

\begin{abstract}
Tritium produced in molten 2LiF--BeF$_2$ (FLiBe) may dissolve as T$_2$ or as TF, which differ
greatly in solubility. Permeation experiments give a solubility close to that of HF and a flux
proportional to the square root of the T$_2$ pressure, which has usually been taken as evidence of
dissociation. We have developed a fixed-charge classical potential for FLiBe containing atomic
tritium, T$_2$, TF, the hydride ion T$^-$ and helium, fitted only to periodic DFT calculations for
hydrogen. The nuclei are classical, so the solubilities apply to all hydrogen isotopes. The host
reproduces the measured density and neutron structure factor but is too fluid close to the
liquidus. The calculated solubilities of T$_2$ and TF agree with experiment, and neither molecule
dissociates in the melt. Because TF is much more soluble, it holds most of the dissolved tritium
unless the salt is strongly reducing, and at constant redox potential its concentration varies
with the square root of the T$_2$ pressure, which accounts for the permeation results without
dissociation. Helium does not change the solubility of T$_2$: a helium bubble in the melt takes up
T$_2$ as its Henry constant predicts, while TF is enriched at the bubble surface. T$_2$ diffuses somewhat faster than measured, and TF and T$^-$ two to three times
more slowly than T$_2$.
\end{abstract}

\section{Introduction}

Molten 2LiF--BeF$_2$ (FLiBe) is the reference salt for fluoride-salt-cooled high-temperature
reactors and several molten salt reactor designs, and a candidate coolant and breeder for fusion
blankets~\cite{Romatoski2017,Fukada2013}. Tritium is produced in the salt by neutron reactions with
$^6$Li and $^7$Li, predominantly as TF~\cite{Fukada2013}. Hydrogen isotopes permeate rapidly
through structural alloys at operating temperature, so the inventory retained in the salt and the
fraction lost through heat-exchanger walls are governed by the solubility and diffusivity of each
tritium species and by the equilibria between them.

These properties are poorly constrained. Malinauskas and Richardson measured the solubility of
H$_2$ and D$_2$ in Li$_2$BeF$_4$ by saturation and stripping and found it to be low and to increase
with temperature~\cite{Malinauskas1974}. HF and DF are more than a hundred times more soluble, and
their solubility decreases with temperature~\cite{Field1967}. In the permeation experiments of
Calderoni \textit{et al.}, T$_2$ entered the salt through a nickel membrane and was collected as
HT in a hydrogen-containing sweep gas; the flux was proportional to the square root of the T$_2$
pressure, and the derived solubility was close to that of HF but increased with
temperature~\cite{Calderoni2008}. The authors concluded that tritium is transported in an ionic
form, while noting that the sign of the solution enthalpy did not support this. Reported
diffusivities span about an order of magnitude, with activation energies between about 30 and
180~kJ/mol~\cite{Calderoni2008,Oishi1989,Anderl2004,Carotti2021}, in part because impurity levels
and redox conditions were not controlled. The redox state also determines the ratio of TF to
T$_2$, and metallic Be is added to FLiBe to reduce TF through Be + 2TF $\rightarrow$ BeF$_2$ +
T$_2$~\cite{Fukada2013}. A third form has also been reported. About 15\% of the tritium in
irradiated FLiBe was found in the $-1$ oxidation state, and after LiH additions electrochemical
measurements detected zero-valent hydrogen at concentrations four orders of magnitude above the
H$_2$ solubility, which was attributed to a BeH$_2$ intermediate~\cite{Carotti2021}.

Ab initio molecular dynamics (AIMD) by Lam \textit{et al.} showed that H$^0$ remains molecular in
FLiBe, whereas H$^+$ binds to fluoride as HF, HF$_2^-$ and complexes with BeF$_4^{2-}$, and that
H$^0$ diffuses three to four times faster than H$^+$~\cite{Lam2021}. Tritium migration had been
studied earlier by AIMD~\cite{Klix2006}, and Galashev \textit{et al.} proposed TF and two forms of
[TF$_2$]$^-$, with equal and unequal T--F distances, from a picosecond trajectory of a small cell
with tritium substituted for Li~\cite{Galashev2023}. Simulations of this size describe the local
structure, but cannot give converged diffusion coefficients of a single solute or solvation free
energies over a range of temperatures.

For the pure salt, rigid-ion Born--Mayer--Huggins models of the Tosi--Fumi type are inexpensive
but describe the structure of BeF$_2$-containing melts poorly~\cite{Smith2020}, whereas the
polarizable ion model of Madden, Salanne and co-workers reproduces the structure and transport of
LiF--BeF$_2$ mixtures~\cite{Salanne2006,Heaton2006}. Neural-network, deep and moment-tensor
potentials trained on DFT agree closely with DFT for the pure
salt~\cite{LamNN2021,Rodriguez2021,Attarian2023}, and the moment-tensor work showed that the density
and transport properties depend on whether dispersion is included in the reference
data~\cite{Attarian2023}. None of these models contains tritium. A machine-learned potential for a
dilute solute needs training data that cover its rare local environments, and it is expensive for
the long single-tracer trajectories required for diffusion.

The aim of this work is to establish which species carries tritium in FLiBe and whether the
permeation results can be explained without dissociation. This requires Henry constants and
diffusion coefficients for each species from one model over the whole temperature range. We
retain a rigid-ion description of the host and add atomic tritium (\Tz), \Tt, TF with a fixed T--F
bond, and the hydride ion (\Tm), which takes the place of a fluoride, as separate species, each through a small set of short-range functions
fitted to DFT. Helium, which is produced together with tritium in the $^6$Li reaction and is used
as a sparge gas, is added in the same way and is used to simulate a helium bubble in the salt, as
in the LIBRA experiment~\cite{Ferry2023}. The fixed T--F bond is checked against DFT calculations
of hydrogen shared between two fluorides. The DFT reference data are for hydrogen. With classical nuclei the solubilities do not depend
on the isotope; the MD uses the tritium mass, and we write the species with T. No solubility or
transport data enter the fit, and since
the potential contains only tabulated pair interactions, Coulomb sums and Morse bonds, it runs in
standard LAMMPS~\cite{Thompson2022}.

\section{Model and methods}

\subsection{Overview of the model}\label{sec:overview}

\begin{figure}[!htbp]
\centering
\includegraphics[width=\textwidth]{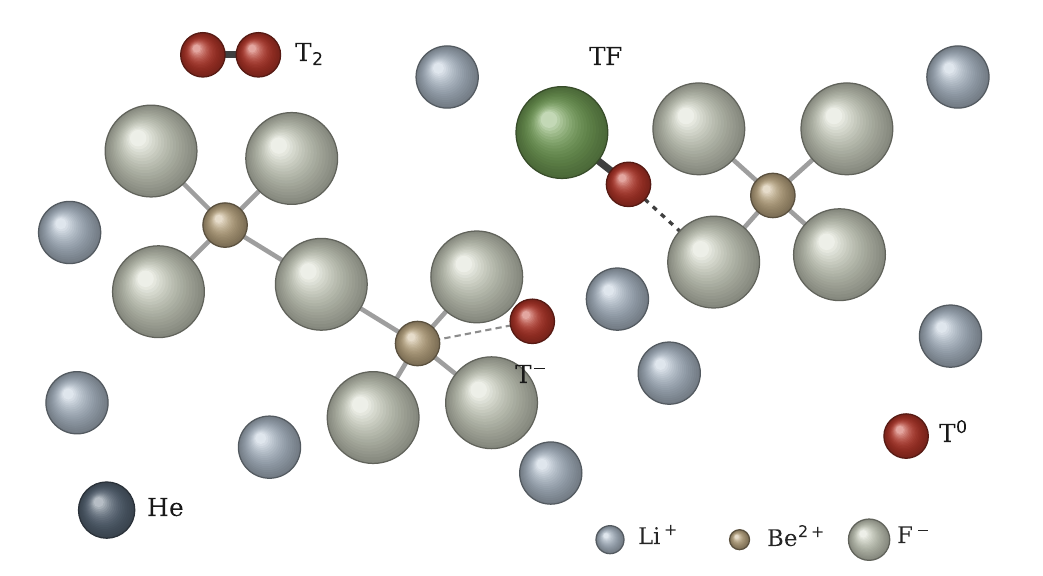}
\caption{Schematic of the dissolved species, each drawn in the environment found in the
simulations. TF (fluoride in green) is hydrogen-bonded to a host fluoride, \Tm{} is next to
Be, and \Tz, \Tt\ and He occupy voids between BeF$_4$ units. The host ions carry scaled formal
charges, T and F of TF carry $\pm q$, \Tm{} carries the charge of the fluoride it replaces,
and \Tz, \Tt\ and He are neutral.}
\label{fig:model}
\end{figure}

The potential energy of FLiBe containing tritium is written as
\begin{equation}
U = U_\mathrm{Coul} + \sum_{i<j\in\text{host}} V_{ab}(r_{ij})
  + \sum_{\alpha} U_\alpha
  + \sum_{\alpha<\beta} U_{\alpha\beta},
\label{eq:total}
\end{equation}
where $U_\mathrm{Coul}$ is the Coulomb energy of all charges, $V_{ab}$ the short-range interaction
between host atoms $i$ and $j$ of species $a$ and $b$, $U_\alpha$ the short-range and bond energy of
tritium species $\alpha$ in the melt, and $U_{\alpha\beta}$ the interaction between two tritium
species. Helium is treated as one more species $\alpha$. All terms are pair functions, Coulomb sums
or Morse bonds (Fig.~\ref{fig:model}). They are listed in Table~\ref{tab:terms}, and the full
expression for each species is given in the Appendix. Atoms are labelled $i$ and $j$, and species
$a$ and $b$ (or $X$ and $Y$).

\begin{table}[!htbp]
\caption{Terms of the model and the reference data used for each.}
\label{tab:terms}
\begin{ruledtabular}\centering
\small
\begin{tabular}{lll}
\hline\hline
interaction & form & reference \\
\hline
host Li$^+$, Be$^{2+}$, F$^-$ & Coulomb + Eq.~(\ref{eq:bmh}) & PBE-D3, $P = 0$ \\
cation--cation correction & table, Eq.~(\ref{eq:ibi}) & NNMD $g(r)$ \\
\Tz--host & Eq.~(\ref{eq:rbf}) & PBE \\
\Tt--host & Morse + Eq.~(\ref{eq:rbf}) & PBE \\
TF--host & Morse + $\pm q$ + Eq.~(\ref{eq:rbf}) & PBE-D3 \\
\Tm--host & $-0.908$\,e + Eq.~(\ref{eq:rbf}) & PBE \\
tritium--tritium & Eq.~(\ref{eq:tt}) & PBE-D3 pairs \\
He--host & Eq.~(\ref{eq:rbf}) & PBE-D3 \\
He--tritium & Eq.~(\ref{eq:tt}) & PBE-D3 pairs \\
He--He & HFD-B & ref~\citenum{Aziz1987} \\
\hline\hline
\end{tabular}
\end{ruledtabular}
\end{table}

The host was fitted first and then kept fixed. Each tritium species was fitted to differences
between DFT energies with and without the species in the same host configuration, so that the
error of the host model does not enter its functions.

\subsection{Host}\label{sec:host}

The ions carry fixed charges of 0.908 times their formal charges. In addition to the Coulomb
energy, two ions interact through
\begin{equation}
V_{ab}(r) = A_{ab} e^{-B_{ab} r} - f_6(B_{ab}r)\frac{C_{6,ab}}{r^6} - f_8(B_{ab}r)\frac{C_{8,ab}}{r^8},
\label{eq:bmh}
\end{equation}
where $f_n(x) = 1 - e^{-x}\sum_{k=0}^{n} x^k/k!$ are the Tang--Toennies damping
functions~\cite{TangToennies1984}.

The parameters were fitted to the forces and energies of PBE-D3 melt configurations, with the
pressure at the experimental density,
\begin{equation}
P = \frac{N k_B T}{V} + \frac{E_\mathrm{Coul}}{3V} - \frac{1}{3V}\sum_{i<j} r_{ij} V_{ab}'(r_{ij}),
\label{eq:pressure}
\end{equation}
added as a further residual at each temperature. A fit to PBE forces alone gives a melt that is
much too light. Since the pressure has to be evaluated on configurations generated by the model
being fitted, the fit was repeated with MD of each new parameter set until the predicted and
simulated pressures agreed. The resulting parameters are listed in Table~\ref{tab:host}, and the
sequence of fits is described in the Supporting Information.

\begin{table}[!htbp]
\caption{Host parameters of Eq.~(\ref{eq:bmh}).$^a$}
\label{tab:host}
\begin{ruledtabular}\centering
\begin{tabular}{lrrrr}
\hline\hline
pair & $A$ (eV) & $B$ (\AA$^{-1}$) & $C_6$ (eV\,\AA$^6$) & $C_8$ (eV\,\AA$^8$) \\
\hline
Li--Li & 4004.9 & 8.56 & 0     & 0 \\
Be--Li & 0      & 8.62 & 1.00  & 0.17 \\
Li--F  & 829.3  & 4.17 & 0.22  & 0.02 \\
Be--Be & 2098.3 & 8.29 & 0.11  & 0.02 \\
Be--F  & 844.4  & 4.23 & 0     & 0 \\
F--F   & 431.6  & 2.89 & 82.18 & 7.97 \\
\hline\hline
\end{tabular}
\end{ruledtabular}
{\par\vspace{2pt}\footnotesize\raggedright $^a$Before the cation--cation correction. Full precision in the released files.\par}
\end{table}

The 84-atom cells contain little information on correlations between cations, and the fitted model
placed Li and Be correctly around F but not around each other. Its Be--Be and Be--Li partial
distribution functions differed from DFT, and corner-sharing BeF$_4$ units were about twice as
frequent. A tabulated correction $\Delta V_{ab}(r)$ was therefore added to the Li--Li, Be--Li and
Be--Be interactions. It was obtained by iterative Boltzmann inversion~\cite{Reith2003} against the
partial distribution functions of the PBE-D3 neural-network MD (NNMD) of Fayfar \textit{et al.} at
973~K~\cite{Fayfar2024},
\begin{equation}
\Delta V_{ab}^{(k+1)}(r) = \Delta V_{ab}^{(k)}(r) + \alpha\,k_BT\ln\frac{g_{ab}^{(k)}(r)}{g_{ab}^\mathrm{ref}(r)},
\label{eq:ibi}
\end{equation}
where $\alpha$ is a damping factor. Each iteration was a short NVT run at the experimental density,
and a small linear term in the Li--Li correction kept the pressure close to zero. The Li--F, Be--F
and F--F interactions were left unchanged. The three cation--cation distribution functions
converged to the NNMD result within about twenty iterations, and the correction averaged over the
last few iterations is used. Although no forces were used to construct it, the correction reduced
the force error on held-out PBE-D3 configurations, mainly on Li, and left the energy error almost
unchanged. The settings are given in the Supporting Information.

\subsection{Tritium--host interactions}\label{sec:tspecies}

The interactions of the tritium species with the host, beyond the host pair terms, are expanded in
Gaussian radial basis functions,
\begin{equation}
U_{XY}(r) = \frac{1}{2}\left[1 + \cos\frac{\pi r}{r_c}\right]\sum_k c_k\,
\exp\!\left[-\frac{(r - r_k)^2}{2\sigma^2}\right],
\label{eq:rbf}
\end{equation}
with the cut-off $r_c$, width $\sigma$ and centre spacing given in Table~\ref{tab:settings}. The
energy is linear in the coefficients $c_k$, so each fit is a linear least-squares problem.

\Tz\ is a neutral atom with functions to Li, Be and F fitted to PBE insertion energies, and \Tt\ is
a Morse-bonded molecule whose atoms interact with the host through functions of the same form, also
fitted to PBE. In TF the tritium carries $+q$ and its fluoride, \Fb, carries $-q$. The T--\Fb\
bond is a Morse function, and the direct T--\Fb\ pair interaction is excluded. \Fb\ keeps the
interactions of a host fluoride together with correction functions to Li, Be and F, and the
tritium has its own functions to Li, Be and F. Because the Coulomb energy is quadratic in $q$, the
reference electrostatics were computed for three charges and the fit was repeated on a grid of $q$;
$q$ and the ridge parameter were chosen from the error on held-out configurations. \Tm{}
replaces a fluoride of the melt and carries the same charge, so that a cell containing \Tm{}
remains neutral, and it has functions to Li, Be and F.

Below the shortest distance $r_g$ sampled in the training data, each fitted function is replaced
by a continuation in which the sum of the function and the Coulomb term equals its linear
extrapolation from $r_g$ plus $B[e^{x} - 1 - x]$, with $x = (r_g - r)/\rho$ and the values of $B$
and $\rho$ in Table~\ref{tab:settings}. The value and slope are continuous at $r_g$, and the
continuation is repulsive whatever the slope of the fitted function at $r_g$.

\subsection{Interactions between tritium species}\label{sec:tt}

When several tritium species share a cell, the neutral species have to be kept from overlapping
each other and TF; without additional terms, \Tz\ atoms in cells containing several of them were
often found on top of each other (Table~\ref{tab:multi}). Site--site functions
\begin{equation}
u(r) = A e^{-br} - f_6(br)\,\frac{C_6}{r^6}
\label{eq:tt}
\end{equation}
were therefore added between \Tz\ and \Tz, between \Tz\ and each atom of \Tt, between the atoms of
two \Tt\ molecules, and between \Tz\ or each atom of \Tt\ and the T and \Fb\ of TF. They were
fitted to PBE-D3 interaction energies of gas-phase pairs. Pairs of \Tz\ atoms were computed in the
triplet state, so that they repel and cannot recombine. Two sites per molecule cannot reproduce the
orientation dependence of the DFT energies, but for \Tt--\Tt\ this dependence is weak, and the
\Tt--\Tt\ and \Tz--\Tt\ functions were fitted to the orientational average, which they reproduce
closely. For \Tt\ near TF the fit was weighted towards the repulsive wall. As a result the shallow
well of a \Tt\ lying across the tritium of TF, which arises from the TF dipole and the \Tt\
quadrupole, is too shallow in the model, although it is already below $k_BT$ in DFT. The curves are
shown in the Supporting Information, and the functions are tested in cells containing many copies
of one species in the section \textit{Several tritium species in one cell}.

Two TF molecules interact through their charges, through the host F--F function between the
fluorides and through the T--F function between the tritium of one and the fluoride of the other,
in the same way as TF interacts with a host fluoride. These terms were not refitted. In the gas
phase the fitted charge gives TF a larger dipole than that of HF and binds the dimer more than
twice as strongly as DFT, but in the melt the same charges reproduce the embedding of TF, and the
interaction between two molecules is screened by the ions. There is no useful gas-phase reference
for \Tm{}. It interacts with \Fb\ and with other \Tm{} ions through the \Tm--F function, with
the tritium of TF through the \Tm--Li function, and with \Tz\ and \Tt\ through their functions to F.
None of these terms allows reactions: two \Tz\ cannot form \Tt, and
$\mathrm{T^- + TF \to T_2 + F^-}$ does not occur.

\subsection{Helium}\label{sec:he}

Helium is a closed-shell neutral atom and is treated in the same way as \Tz. Its functions to Li,
Be and F, of the form of Eq.~(\ref{eq:rbf}), were fitted to PBE-D3 insertion energies and forces
of He in the melt, with the He placed at the centres of the largest cavities, at random positions
and pressed against host ions. The cavity and random sites, which determine the solubility, were
given the largest weight, and the contact configurations fix the repulsive walls. He--He
interactions are described by the HFD-B function of Aziz \textit{et al.}~\cite{Aziz1987}. He
interacts with \Tz, with each atom of \Tt\ and with the T and F of TF through functions of the form
of Eq.~(\ref{eq:tt}) fitted to gas-phase PBE-D3 pairs, and with \Tm{} through its function to
F. The fit errors and parameters, and a comparison of the He--\Tz\ interaction with the
calculations of Tirumala \textit{et al.}~\cite{Tirumala2026} and with other available potentials~\cite{yang2020first,chen2022md}, are given in the Supporting
Information.

\subsection{DFT reference data}
\label{sec:dft}

All DFT calculations were carried out with CP2K~\cite{Kuhne2020} using the PBE
functional~\cite{PBE1996}, Goedecker--Teter--Hutter pseudopotentials~\cite{Goedecker1996},
TZV2P MOLOPT basis sets~\cite{VandeVondele2007}, a plane-wave cut-off of 1200~Ry and the $\Gamma$
point. D3 dispersion with Becke--Johnson damping~\cite{Grimme2010,Grimme2011} was included for all
data except the \Tz, \Tt\ and \Tm{} sets. The melt cells contain 12 Li$_2$BeF$_4$ units at the
experimental density together with the hydrogen species, and were taken from classical MD over the
whole temperature range. Atomic charges were obtained from the Hirshfeld
partition~\cite{Hirshfeld1977} and, for \Tm{}, from maximally localized Wannier
functions~\cite{Marzari2012}. The data sets are listed in Table~S1.

The gas-phase pairs used for the tritium--tritium interactions were computed with PBE-D3 in a
periodic cell, spin-polarized for open shells (\Tz--\Tz\ as a triplet and pairs containing one \Tz\
as doublets), relative to the separated partners in the same cell. They cover \Tz--\Tz\ separations
and several orientations of the \Tz--\Tt, \Tt--\Tt, \Tz--TF, \Tt--TF and TF--TF pairs, with the TF
bond at its model equilibrium length.

\subsection{Fitting targets and sampling}

For TF the energy target was the embedding energy
\begin{equation}
\Delta E = E(\text{host} + \mathrm{TF}) - E(\text{host}) - E_\mathrm{TF}^\mathrm{gas}(r_\mathrm{TF}),
\label{eq:embed}
\end{equation}
with identical host atoms in the first two terms and the isolated molecule at the same bond length.
The force targets were the forces on T and \Fb\ minus the bond force of the isolated molecule, and
the forces induced on the host atoms, $\mathbf{F}_i(\text{host}+\mathrm{TF}) -
\mathbf{F}_i(\text{host})$. No tritium function contains the cation--cation correction of the
host, so the correction leaves the tritium functions unchanged. The D3 contribution to
Eq.~(\ref{eq:embed}), evaluated on paired configurations, is nearly constant and was added to the
targets so that TF is described at the same level as the host. For \Tm{} the targets were
the substitution energies $E(\text{F}_i \rightarrow \text{T}) - E(\text{host})$, with a single
constant for the F/T reference, and the forces after removal of the host-model error of the parent
configuration.

The TF training set combines MD frames from earlier TF models, bond-stretch scans, isolated
molecules, configurations visited in MD by a first refitted model, and scans that move the tritium
from its fluoride towards a second one. Configurations in which the hydrogen sits on a host fluoride
cannot be described with a fixed bond and were excluded. A fifth of the configurations were held
out for validation. The successive training sets are described in the Supporting Information.

\subsection{Molecular dynamics}

MD simulations were run with LAMMPS~\cite{Thompson2022} using Nos\'e--Hoover thermostats and PPPM
electrostatics, in cubic cells at the experimental density~\cite{Romatoski2017} containing a single
tritium species. Cell sizes, time step and run lengths are given in Table~\ref{tab:settings}. Shear
viscosities were calculated from the Green--Kubo integral of the off-diagonal pressure tensor, and
electrical conductivities from the mean squared displacement of the total dipole. Diffusion
coefficients were obtained from the mean squared displacement of the tritium, or of the centre of
mass of \Tt, averaged over all time origins and corrected for drift of the cell, in four
independent cells per temperature. They were corrected for the finite cell size with the
expression of Yeh and Hummer~\cite{YehHummer2004}, $D_\infty = D_L + \xi k_BT/(6\pi\eta L)$, using
the viscosity of the model, and the correction was checked with larger cells.

\begin{table}[!htbp]
\caption{Computational settings.}
\label{tab:settings}
\begin{ruledtabular}\centering
\small
\begin{tabular}{ll}
\hline\hline
DFT & CP2K, PBE and PBE-D3(BJ) \\
basis, cut-off & GTH, TZV2P-MOLOPT, 1200~Ry, $\Gamma$ \\
DFT melt cell & 84 atoms, $L \approx 10.1$~\AA \\
gas-phase pairs & 14~\AA\ cubic cell \\
host charges (e) & Li $+0.908$, Be $+1.816$, F $-0.908$ \\
host cut-off & 9.37~\AA \\
Eq.~(\ref{eq:rbf}): $r_c$, $\sigma$, spacing & 5, 0.35, 0.25~\AA \\
continuation: $B$, $\rho$ & 5~eV, 0.1~\AA \\
training sets & TF 690, He 252, pairs 338 configurations \\
time step & 0.25~fs \\
PPPM accuracy & $10^{-5}$ \\
production cells & 756 atoms, 4 $\times$ 1~ns per temperature \\
size check & 1792 atoms, 973 and 1273~K \\
MSD fit window & 20--100~ps \\
test insertion & $2\times10^4$ per frame, frames every 2~ps \\
TI & 12 + 8 Gauss--Legendre points, 40 + 200~ps per window \\
several species & 3500 atoms, 4 or 16 copies, 500~ps \\
He bubble & $\approx$20\,000 atoms, $R = 12$~\AA, 150 He, 16 \Tt, 16 TF, 2~ns \\
\hline\hline
\end{tabular}
\end{ruledtabular}
\end{table}

\subsection{Solubility}\label{sec:sol_methods}

For a dilute solute in equilibrium with the same species in an ideal gas, the kinetic and
intramolecular contributions to the chemical potential are the same in both phases in classical
statistical mechanics, and the ratio of number densities is the Ostwald coefficient
\begin{equation}
L = \frac{\rho_\mathrm{salt}}{\rho_\mathrm{gas}} = e^{-\beta\muex},
\label{eq:ostwald}
\end{equation}
where $\beta = 1/k_BT$ and \muex\ is the excess chemical potential of the solute in the melt. With
$c_\mathrm{gas} = p/RT$, the Henry constant defined by $c_\mathrm{salt} = K_H\,p$ is
\begin{equation}
K_H = \frac{L}{RT} = \frac{e^{-\beta\muex}}{RT}\quad(\mathrm{mol\,m^{-3}\,Pa^{-1}}),
\label{eq:KH}
\end{equation}
and the solution enthalpy, $\Delta H_s = -R\,\mathrm{d}\ln K_H/\mathrm{d}(1/T)$, was obtained
from a linear fit of $\ln K_H$ against $1/T$. The classical \muex\ does not depend on the nuclear
masses, so the results apply to all hydrogen isotopes and can be compared directly with
measurements for H$_2$ and HF.

For the neutral species \muex\ was computed by test-particle insertion~\cite{Widom1963},
\begin{equation}
\muex = -k_BT\ln\left\langle e^{-\beta\,\Delta U(\mathbf{r},\Omega)}\right\rangle_{N},
\label{eq:widom}
\end{equation}
where $\Delta U$ is the interaction energy of a test \Tz\ or \Tt\ at a random position $\mathbf{r}$,
and for \Tt\ a random orientation $\Omega$, in configurations of the pure melt at the experimental
density. \Tt\ was inserted as a rigid molecule at its gas-phase bond length, so that its
intramolecular energy cancels. Configurations were taken from four independent cells per
temperature after removing the tritium of the stored runs, and the errors are standard errors over
the cells.

Test insertion converges poorly for TF because of its charges, and its \muex\ was obtained instead
by thermodynamic integration~\cite{Kirkwood1935} in two stages, first switching on the short-range
TF--host interactions of an uncharged molecule (stage B) and then its charges (stage A),
\begin{equation}
\muex = \Delta A_B + \Delta A_A = \int_0^1\left\langle\frac{\partial U}{\partial\lambda}\right\rangle_{\lambda}^{(B)}\!\mathrm{d}\lambda
+ \int_0^1\left\langle\frac{\partial U}{\partial\lambda}\right\rangle_{\lambda}^{(A)}\!\mathrm{d}\lambda .
\label{eq:ti}
\end{equation}
Here $\lambda$ is a coupling parameter that switches the interaction of TF with the melt on
gradually, from no interaction at $\lambda = 0$ to full interaction at $\lambda = 1$, and
$\langle\cdots\rangle_\lambda$ is an average over MD at fixed $\lambda$. In stage B the short-range energy $U_\mathrm{sr}$ is scaled by $\lambda^4$, which removes the
end-point divergence and gives $\partial U/\partial\lambda = 4\lambda^3 U_\mathrm{sr}$. In stage A
the charges are $\pm\lambda q$ and the energy is quadratic in $\lambda$, $U(\lambda) = U_0 +
a\lambda + b\lambda^2$, so that $\partial U/\partial\lambda = a + 2b\lambda$, with $a$ and $b$
obtained by re-evaluating each stored frame at $\lambda = 0$ and $\pm 1$. Both integrals were
evaluated by Gauss--Legendre quadrature. At $\lambda = 0$ in stage B the molecule keeps its bond but
does not interact with the melt, which is the ideal-gas reference of Eq.~(\ref{eq:ostwald}). The
intramolecular T--F pair is excluded from both the short-range and the Coulomb terms, and no
self-energy correction is needed. Errors were estimated from block averages in each window.

TF was fitted to PBE-D3 data, so its \muex\ is at the PBE-D3 level. The \Tz\ and \Tt\ functions
were fitted to PBE, and their \muex\ was corrected to PBE-D3 by free-energy perturbation,
\begin{equation}
\Delta\muex^\mathrm{D3} = -k_BT\ln\left\langle e^{-\beta(\Delta E_\mathrm{D3} - \Delta E_\mathrm{PBE})}\right\rangle_{\Delta U},
\label{eq:d3}
\end{equation}
where $\Delta E_\mathrm{PBE}$ and $\Delta E_\mathrm{D3}$ are DFT insertion energies at the same
site, and the average runs over sites drawn with probability $\propto e^{-\beta\Delta U}$ from the
model insertion energies, that is, over the sites that dominate Eq.~(\ref{eq:widom}). For \Tz\ the
correction was computed at three temperatures and interpolated, and for \Tt\ paired PBE and PBE-D3
insertions gave a constant shift. Both corrections are reported in the Supporting Information. The
He functions were fitted directly to PBE-D3, and He was inserted in the same way as \Tz\ without a
correction.

\section{Validation}

\subsection{Against DFT}

\begin{figure}[!htbp]
\centering
\includegraphics[width=\textwidth]{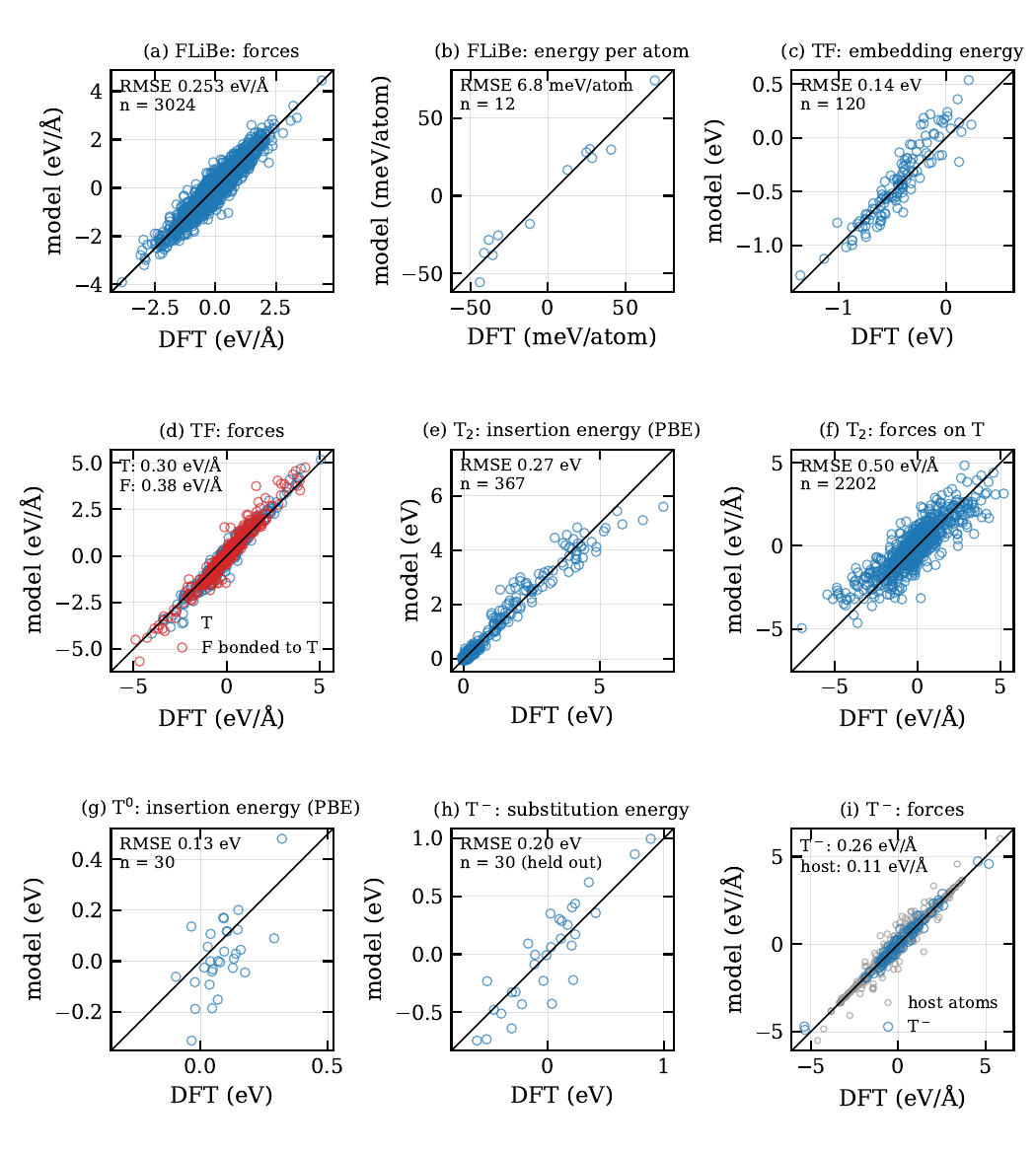}
\caption{Model against DFT. (a) Forces and (b) energies per atom (mean removed) of held-out host
cells (PBE-D3). (c) TF embedding energies, Eq.~(\ref{eq:embed}), for the 120 configurations of the
first data set, evaluated with the complete model; the PBE values include the D3 shift. (d) Forces
on T and on its fluoride in the same configurations. (e) \Tt\ insertion energies and (f) forces on
T, held out. (g) \Tz\ insertion energies at Boltzmann-sampled sites. (h) \Tm\ substitution energies
(mean removed) and (i) forces on T and on the host atoms, held out.}
\label{fig:parity}
\end{figure}

Figure~\ref{fig:parity} and Table~\ref{tab:dfterr} compare the model with held-out DFT data. The
host improves on the force-only rigid-ion model used as the starting point, and its largest force
error is on Be, which is consistent with the lack of fluoride polarization around a small, highly
charged cation. In both DFT and the model the T--F bond of TF is longer in the melt than in the
isolated molecule. The fitted TF charge is close to the Hirshfeld charge of the hydrogen in the
melt, but the validation error hardly changes over a range of $q$, so the charge is determined only
to about $\pm 0.1$\,e. For \Tt\ most of the error comes from compressed configurations that are
rarely visited in MD.

\begin{table}[!htbp]
\caption{RMS errors against held-out DFT data.}
\label{tab:dfterr}
\begin{ruledtabular}\centering
\begin{tabular}{lccc}
\hline\hline
& energy & $F$ on solute & $F$ on host \\
& (eV) & (eV/\AA) & (eV/\AA) \\
\hline
host & 0.007$^a$ & -- & 0.25$^b$ \\
TF & 0.14 & 0.30, 0.38$^c$ & -- \\
\Tt & 0.27 & 0.50 & -- \\
\Tz & 0.13 & -- & -- \\
\Tm & 0.20 & 0.26 & 0.11 \\
He & 0.01$^d$ & 0.04$^d$ & -- \\
\hline\hline
\end{tabular}
\end{ruledtabular}
{\par\vspace{2pt}\footnotesize\raggedright $^a$Per atom. $^b$Be 0.41. $^c$T and \Fb. $^d$Cavity sites.\par}
\end{table}

\subsection{Molten FLiBe}\label{sec:flibe}

\begin{figure}[!htbp]
\centering
\includegraphics[width=\textwidth]{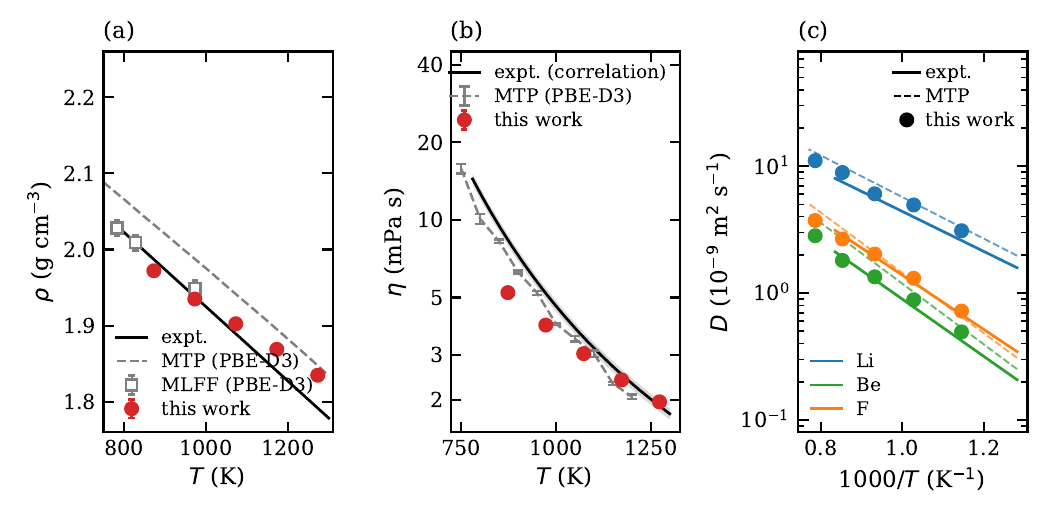}
\caption{Molten FLiBe with the host model (circles). (a) Density at zero pressure, with the
recommended correlation~\cite{Romatoski2017}, the MTP of Ref.~\onlinecite{Attarian2023} and the
machine-learned force field of Ref.~\onlinecite{Mlff2026}, both PBE-D3. (b) Green--Kubo viscosity
with the correlation of Ref.~\onlinecite{Tasidou2019} and its uncertainty band.
(c) Self-diffusion coefficients with Arrhenius fits to the data of Ref.~\onlinecite{Mei2013} (solid)
and to the MTP (dashed).}
\label{fig:host}
\end{figure}

The host is compared with experiment in Tables~\ref{tab:hostval} and \ref{tab:hostact} and in
Fig.~\ref{fig:host}. The density agrees with the recommended correlation~\cite{Romatoski2017} to
within a few percent, but the thermal expansion is too small, so that the model is slightly too
light at 873~K and slightly too dense at 1273~K. The cation--cation correction changes the density
by less than 1\%.

The correction lowers the viscosity. Compared with the correlation of Tasidou \textit{et
al.}~\cite{Tasidou2019}, the model viscosity is close to experiment at high temperature but too low
near the liquidus, and the apparent activation energy is therefore too small. Before the correction
the viscosity was close to experiment, but only because the model contained twice as many
corner-sharing BeF$_4$ units as DFT, and this over-connected network made up for the missing
viscosity. Once the connectivity matches DFT, a rigid-ion model without polarization of the
bridging fluoride cannot reproduce the viscosity. The moment-tensor potential (MTP) of Attarian
\textit{et al.}, fitted to PBE-D3, is close to the correlation at 873~K~\cite{Attarian2023}, which
places the error in the fixed-charge model rather than in the functional.

The self-diffusion coefficients show the same trend. After the finite-size correction the ions
diffuse somewhat faster than measured~\cite{Mei2013}, Be most of all at 873~K, and the activation
energies of the two network-forming ions are too low (Table~\ref{tab:hostact}). The electrical
conductivity lies below the Nernst--Einstein estimate, slightly at 873~K and by about a third at
1273~K.

\begin{table}[!htbp]
\caption{Density, viscosity, conductivity and Be--F coordination of the host.}
\label{tab:hostval}
\begin{ruledtabular}\centering
\begin{tabular}{rcccccc}
\hline\hline
$T$ & \multicolumn{2}{c}{$\rho$ (g/cm$^3$)} & \multicolumn{2}{c}{$\eta$ (mPa\,s)} & $\sigma$ (S/m) & $N_\mathrm{BeF}$ \\
(K) & model & expt.~\cite{Romatoski2017} & model & expt.~\cite{Tasidou2019} & model & model \\
\hline
873  & 1.97 & 1.99 & 5.2 & 8.4 & 217 & 4.02 \\
973  & 1.94 & 1.94 & 3.9 & 5.2 & 283 & 4.03 \\
1073 & 1.90 & 1.89 & 3.0 & 3.5 & 335 & 4.05 \\
1173 & 1.87 & 1.84 & 2.4 & 2.5 & 383 & 4.03 \\
1273 & 1.84 & 1.79 & 2.0 & 1.9 & 424 & 4.04 \\
\hline\hline
\end{tabular}
\end{ruledtabular}
\end{table}

\begin{table}[!htbp]
\caption{Activation energies of the host (kJ/mol), 873--1273~K.}
\label{tab:hostact}
\begin{ruledtabular}\centering
\begin{tabular}{lcc}
\hline\hline
& model & expt. \\
\hline
$\eta$ & 23 & 34~\cite{Tasidou2019} \\
$D_\mathrm{Li}$ & 29 & 30~\cite{Mei2013} \\
$D_\mathrm{Be}$ & 37 & 43~\cite{Mei2013} \\
$D_\mathrm{F}$ & 36 & 41~\cite{Mei2013} \\
\hline\hline
\end{tabular}
\end{ruledtabular}
\end{table}

Be is fourfold coordinated, with a few percent of Be atoms carrying a fifth fluoride, more at high
temperature. Between a half and a third of the BeF$_4$ units, fewer at higher temperature, are
isolated, and the rest share corners through bridging fluorides, which make up about a tenth of all
fluorides.

\begin{figure}[!htbp]
\centering
\includegraphics[width=0.75\textwidth]{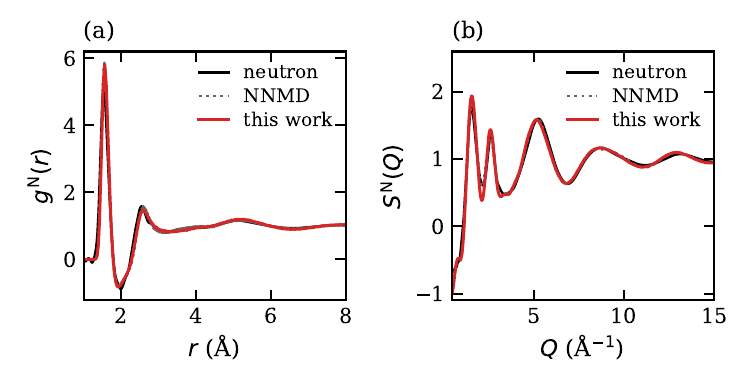}
\caption{(a) Neutron-weighted total $g(r)$ and (b) structure factor of FLiBe at 973~K from this
work and the NNMD of Ref.~\onlinecite{Fayfar2024}, with the neutron measurement at 700~$^\circ$C
($^7$Li-enriched salt) from the same work. Faber--Ziman normalization; the model and NNMD $S(Q)$
are Fourier transforms of the partials to $r = L/2$ without a window.}
\label{fig:scatter}
\end{figure}

Figure~\ref{fig:scatter} compares the neutron-weighted total $g(r)$ and $S(Q)$ with the
measurement of Fayfar \textit{et al.} at 700~$^\circ$C. Apart from the correction of the
cation--cation functions against NNMD, no structural data entered the fit. The model reproduces
the measured $S(Q)$, including the position of the first peak, which is set by cation--cation
correlations and was too high in the uncorrected model (Table~\ref{tab:struct}).

\begin{figure}[!htbp]
\centering
\includegraphics[width=\textwidth]{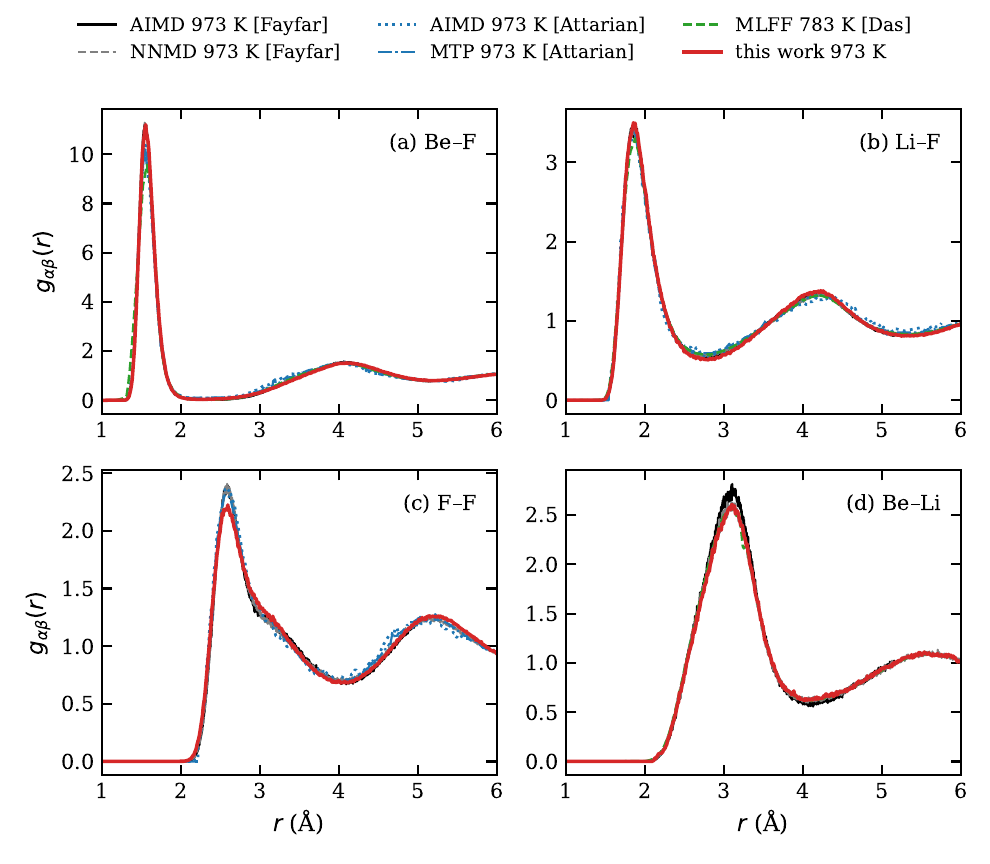}
\caption{Be--F, Li--F, F--F and Be--Li radial distribution functions of molten FLiBe: this work at
973~K (red); AIMD and NNMD at 973~K from Ref.~\onlinecite{Fayfar2024}; AIMD and MTP at 973~K from
Ref.~\onlinecite{Attarian2023}; machine-learned force field at 783~K from Ref.~\onlinecite{Mlff2026}.
All reference simulations are PBE-D3. The curves of Refs.~\onlinecite{Attarian2023} and
\onlinecite{Mlff2026} were digitized from the published figures.}
\label{fig:partials}
\end{figure}

Figure~\ref{fig:partials} and Table~\ref{tab:struct} compare the partial distribution functions at
973~K with PBE-D3 simulations: the AIMD and NNMD of Fayfar \textit{et al.}~\cite{Fayfar2024}, the
AIMD and MTP of Attarian \textit{et al.}~\cite{Attarian2023} and the machine-learned force field of
Das \textit{et al.} at 783~K~\cite{Mlff2026}. The Be--F and Li--F shells agree with DFT, and the
F--F peak lies at the DFT position with a lower maximum. The cation--cation correlations now agree
with DFT, whereas before the correction Be--Be had a sharp peak near 3.1~\AA\ and almost twice the
DFT number of close Be neighbours.

The neutron analysis gives a longer Li--F distance and a Be coordination number below four. Both
come from fits to the total $g(r)$, in which Li--F has a negative weight and overlaps the tail of
the Be--F peak, while the simulations that reproduce the measured $S(Q)$ all give four fluorides
around Be.

\begin{table}[!htbp]
\caption{Structure of molten FLiBe at 973~K.$^a$}
\label{tab:struct}
\begin{ruledtabular}\centering
\small
\begin{tabular}{lcccc}
\hline\hline
& this work & uncorrected & DFT$^b$ & expt. \\
\hline
$r_\mathrm{BeF}$ (\AA) & 1.55 & & 1.54, 1.55 & 1.53,$^c$ 1.58$^d$ \\
$N_\mathrm{BeF}$ & 4.03 & & 3.97--3.99 & $3.7 \pm 0.15$$^c$ \\
$r_\mathrm{LiF}$ (\AA) & 1.86 & & 1.87 & 1.92--1.95 \\
$N_\mathrm{LiF}$ ($<$2.75~\AA) & 4.52 & & 4.50, 4.52 & \\
$r_\mathrm{FF}$ (\AA) & 2.59 & & 2.59 & \\
$g_\mathrm{FF}^\mathrm{max}$ & 2.23 & & 2.39--2.41 & \\
$r_\mathrm{BeLi}$ (\AA) & 3.10 & & 3.11 & \\
$N_\mathrm{BeLi}$ ($<$3.0~\AA) & 2.25 & & 2.3 & \\
$N_\mathrm{BeLi}$ ($<$4.1~\AA) & 7.6 & & 7.6--7.8 & \\
$r_\mathrm{BeBe}$ (\AA) & 5.04 & 3.07 & 4.97, 5.02 & \\
$N_\mathrm{BeBe}$ ($<$3.5~\AA) & 0.72 & 1.3 & 0.6--0.7 & \\
$r_\mathrm{LiLi}$ (\AA) & 3.09 & & 3.09 & \\
$g_\mathrm{LiLi}^\mathrm{max}$ & 1.60 & & 1.59 & \\
$S(Q)$ peaks (\AA$^{-1}$) & 1.58, 2.60, 5.16 & 1.73 & & 1.56, 2.64, 5.26$^c$ \\
\hline\hline
\end{tabular}
\end{ruledtabular}
{\par\vspace{2pt}\footnotesize\raggedright $^a$$r$: main peak; $N$: coordination number to the first minimum or to the distance given. Uncorrected: before the cation--cation correction. $^b$NNMD and AIMD.\cite{Fayfar2024} $^c$Neutron.\cite{Fayfar2024} $^d$X-ray.\cite{Vaslow1973}\par}
\end{table}

The F--F dispersion coefficient (Table~\ref{tab:host}) is large for a fluoride. The pressure
condition acts mainly on this parameter, which in a rigid-ion model takes the place of the
polarization of F$^-$, a large part of the cohesion of BeF$_2$-rich melts in polarizable
models~\cite{Salanne2006}. It should be regarded as an effective parameter of this model.

\section{Tritium species in the melt}

\subsection{TF and shared F--T$\cdots$F configurations}
\label{sec:tf_struct}

\begin{figure}[!htbp]
\centering
\includegraphics[width=\textwidth]{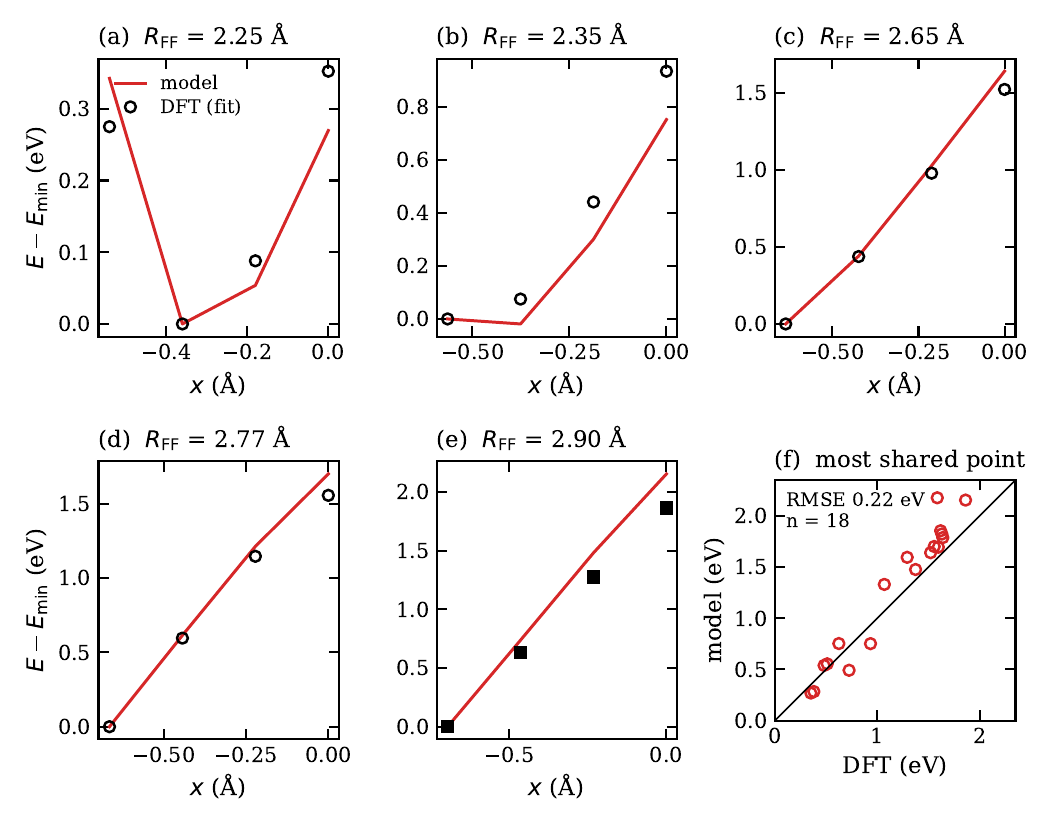}
\caption{Hydrogen between its fluoride and a second fluoride in the melt. (a)--(e) Energy along the
transfer coordinate $x$ for five F$\cdots$F distances $R_\mathrm{FF}$, relative to the lowest DFT
point of each scan: PBE-D3 (symbols; filled, held out of the fit) and the model with the bond kept on
the original fluoride (line). (f) Energy of the most shared configuration of every scan, model
against DFT.}
\label{fig:ftf}
\end{figure}

A fixed T--F bond is justified only if the tritium of TF does not transfer between fluorides. AIMD
finds HF and HF$_2^-$ in FLiBe~\cite{Lam2021}, so the tritium does interact with a second fluoride.
In every F--T--F scan in the melt, however, the energy minimum stays on the side of the original
fluoride, and moving the hydrogen onto the second fluoride costs 2--3~eV (Fig.~\ref{fig:ftf}). The
symmetric, shared position is least unfavourable at F$\cdots$F distances close to the bifluoride
geometry, but even there it lies more than three times $k_BT$ above the minimum at 1273~K, so
transfer does not contribute to transport. The model reproduces the energy of the most shared point
of each scan (Fig.~\ref{fig:ftf}f).

\begin{table}[!htbp]
\caption{Surroundings of the tritium species at 973~K.$^a$}
\label{tab:env}
\begin{ruledtabular}\centering
\begin{tabular}{lcccc}
\hline\hline
& \Tz & \Tt & TF & \Tm \\
\hline
bond length (\AA) & -- & 0.76 & 1.00 & -- \\
$r_\mathrm{TF}$ (\AA) & 2.6 & 3.0 & 1.43 & -- \\
$N_\mathrm{F}$ ($<$2~\AA) & 0.14 & 0.04 & 0.96 & -- \\
$r_\mathrm{TLi}$ (\AA) & 2.5 & 3.7$^b$ & -- & 1.95--2.11 \\
$N_\mathrm{Li}$ ($<$3~\AA) & 2.9 & -- & 1.7--2.0$^c$ & -- \\
$N_\mathrm{Be}$ ($<$3~\AA) & 0.6 & 0.6 & 0.6 & -- \\
$r_\mathrm{TBe}$ (\AA) & -- & -- & -- & 1.72, 2.27 \\
\hline\hline
\end{tabular}
\end{ruledtabular}
{\par\vspace{2pt}\footnotesize\raggedright $^a$$r$: first peak; $N$: neighbours within the distance given. The bound fluoride of TF and the second atom of \Tt\ are excluded. $^b$Broad maximum. $^c$Around \Fb. 19, 44 and 37\% of \Tm{} ions have 1, 2 and 3 or more Be within 2.5~\AA.\par}
\end{table}

In MD the TF molecule forms a hydrogen bond to a neighbouring fluoride, and the T--\Fb\ bond
lengthens (Table~\ref{tab:env}). In about 1\% of the frames the tritium is briefly closer to a host
fluoride than to its own. This hydrogen-bonded unit corresponds to the asymmetric [F--T$\cdots$F]$^-$
of Galashev \textit{et al.}~\cite{Galashev2023}. Their symmetric form, with two equal bonds,
corresponds to the shared point of our scans and can at most be short-lived between 873 and 1273~K;
a picosecond trajectory of a small cell samples too few of these events to establish their
populations.

A first refitted TF model, trained without configurations of this kind, pulled the tritium onto a
host fluoride in MD despite a small held-out error. For this reason configurations visited by the
model itself were added to the training data (Supporting Information). The T--F(host)
distribution of the final model at 973~K is shown in Fig.~\ref{fig:tsites}(b).

In the production runs the fluoride of TF does not enter the first shell of Be and has one or two
Li neighbours. AIMD has been interpreted as showing hydrogen attached to BeF$_4$
units~\cite{Lam2021,Galashev2023}, and since the TF model was trained largely on configurations it
generated itself, this result was checked with DFT. The hydrogen was placed on a terminal fluoride of
a BeF$_4$ unit in twelve configurations taken from six melt cells, which were then relaxed with
PBE-D3 (Supporting Information). None kept a Be--F--H arrangement: the fluoride left the Be, the
hydrogen moved to a fluoride not bonded to Be, or it ended up shared between two fluorides. The
relaxed structures lie on average about half an electronvolt above free TF in the same cell. The
model, which cannot transfer the tritium, relaxes all twelve configurations by detaching the
fluoride from the Be. DFT and the model therefore agree that hydrogen does not stay on a fluoride
bonded to Be. The shared F--T--F configurations that DFT also finds cannot be represented by a
non-reactive model, and they are a likely source of the Be-associated hydrogen reported from AIMD.
In relaxed free TF the second fluoride is at about the distance of the hydrogen-bond peak of the
model.

\subsection{The hydride ion \Tm}\label{sec:hydride}

The Hirshfeld charge of hydrogen substituted for a fluoride is about $+0.5$\,e, which would point to
an oxidized species. The same partition, however, gives a similar charge to hydrogen in crystalline
LiH computed with the same settings, so Hirshfeld charges cannot distinguish a hydride from a proton
when the hydrogen is surrounded by Li or Be. Wannier centres can. In LiH the centre of the hydrogen
electron pair lies on the nucleus, and in TF it lies on the bond, roughly halfway to F. For hydrogen
at free, terminal and bridging fluoride sites the centre remains on the hydrogen and moves only
slightly towards Be, even when a Be is very close (Supporting Information). The hydrogen is
therefore a hydride ion, and a full ionic charge is appropriate.

In MD \Tm{} sits between Li and Be (Table~\ref{tab:env}), with the nearest Li at about the
Li--H distance of crystalline LiH. The T--Be distribution has two peaks (Fig.~\ref{fig:tsites}d).
Since $U_\mathrm{T^-Be}$ plus the Coulomb term has a single minimum, the second peak is not an
artefact of the fit but a second Be, and most \Tm{} ions have two or three Be within 2.5~\AA, more
than a bridging fluoride. Carotti \textit{et al.} proposed BeH$_2$ as the soluble form of hydrogen
after LiH addition~\cite{Carotti2021}. In the model \Tm{} is shared between Be atoms as in
BeH$_2$, but as an ion with a fixed charge and without a covalent Be--H bond; the extent of this
sharing in DFT was not examined beyond the training configurations. The instability of \Tm{}
with respect to 2LiH + BeF$_2$ $\rightarrow$ 2LiF + Be + H$_2$ is outside the scope of a
fixed-charge model.

\subsection{\Tt\ and \Tz}

\begin{figure}[!htbp]
\centering
\includegraphics[width=\textwidth]{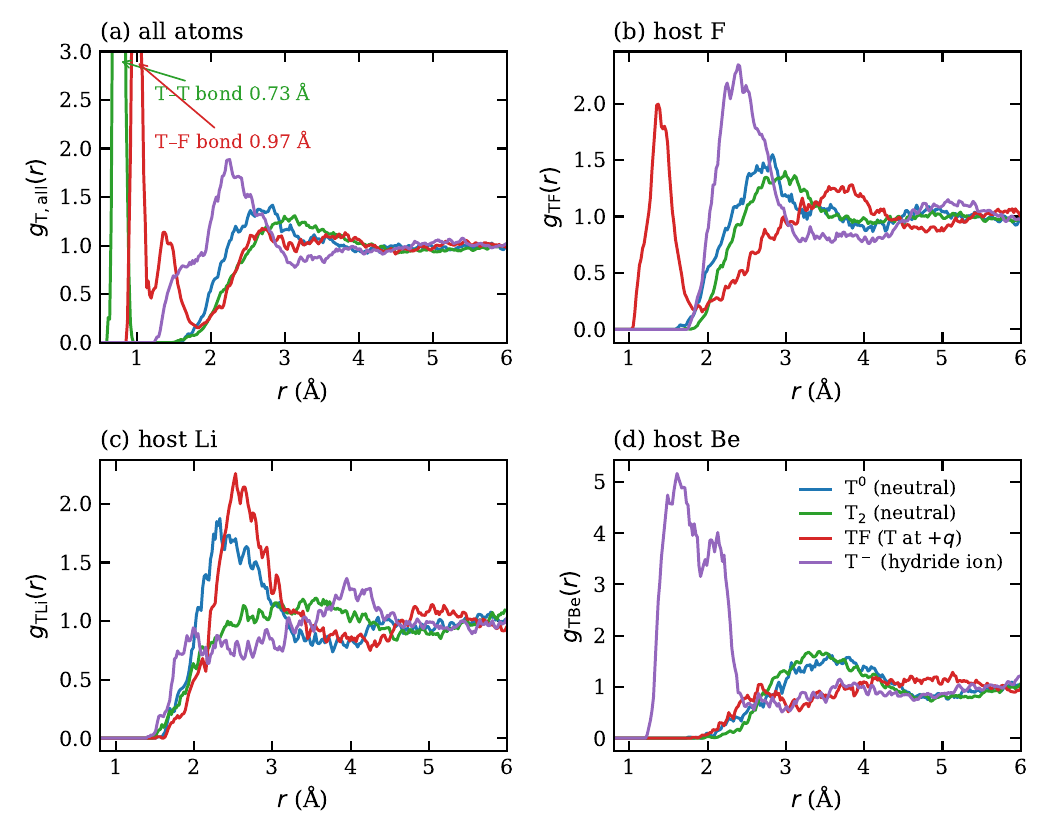}
\caption{Radial distribution functions between the hydrogen of each species and (a) all other atoms,
(b) host F, (c) Li and (d) Be at 973~K. \Tz\ and \Tt\ are neutral, the hydrogen of TF carries $+q$,
and \Tm\ carries $-0.908$\,e. Panel (a) includes the bonded partner, so the T--T bond of \Tt\ near
0.75~\AA\ and the T--F bond of TF near 1.0~\AA\ appear as sharp peaks, cut off at the top of the
axis. In (b)--(d) the bound fluoride of TF and the second atom of \Tt\ are excluded. Single-tracer runs, four per
species.}
\label{fig:tsites}
\end{figure}

\Tt\ remains molecular, in agreement with AIMD~\cite{Lam2021}. Both neutral species occupy voids in
the ionic network and have almost no fluoride within 2~\AA\ (Table~\ref{tab:env}), and in the
stability runs their closest approaches to cations stay outside the short-range continuation of
their functions. \Tz\ has a well-defined Li shell, whereas \Tt\ shows only a broad Li maximum at
larger distance. \Tz, \Tt\ and TF all have fewer than one Be within 3~\AA, consistent with sites at
the surface of the BeF$_4$ units rather than in their first shell. The four species are compared in
Fig.~\ref{fig:tsites}.

\section{Solubility and redox}
\label{sec:sol}

The Henry constants and excess chemical potentials are given in Tables~\ref{tab:sol} and
\ref{tab:muex}. Without the D3 correction the Henry constant of \Tt\ would be several times lower.

\begin{table}[!htbp]
\caption{Henry constants (mol\,m$^{-3}$\,Pa$^{-1}$) and solution enthalpies.$^a$}
\label{tab:sol}
\begin{ruledtabular}\centering
\small
\setlength{\tabcolsep}{4pt}
\begin{tabular}{rcccccc}
\hline\hline
& \multicolumn{3}{c}{this work} & \multicolumn{3}{c}{expt.} \\
$T$ (K) & \Tt & TF & He & H$_2$~\cite{Malinauskas1974} & HF~\cite{Field1967} & He~\cite{Malinauskas1974} \\
\hline
873  & $4.7\times10^{-7}$ & $2.6\times10^{-4}$ & $2.3\times10^{-7}$ & $4\times10^{-7}$ & $1.3\times10^{-4}$ & $8.3\times10^{-7}$  \\
973  & $8.1\times10^{-7}$ & $1.4\times10^{-4}$ & $4.0\times10^{-7}$ & $5\times10^{-7}$ & $9\times10^{-5}$ & $1.1\times10^{-6}$  \\
1073 & $1.3\times10^{-6}$ & $1.1\times10^{-4}$ & $6.3\times10^{-7}$ & -- & $7\times10^{-5}$$^b$ & $1.6\times10^{-6}$  \\
1173 & $1.8\times10^{-6}$ & $1.0\times10^{-4}$ & $9.3\times10^{-7}$ & -- & -- & --  \\
1273 & $2.5\times10^{-6}$ & $1.1\times10^{-4}$ & $1.3\times10^{-6}$ & -- & -- & --  \\
\hline
$\Delta H_s$ (kJ/mol) & $+39 \pm 1$ & $-17 \pm 5$ & $+40$ & $+29$ & $-26$ & $+23$ \\
\hline\hline
\end{tabular}
\end{ruledtabular}
{\par\vspace{2pt}\footnotesize\raggedright $^a$Measured H$_2$ and He: Ostwald coefficients in Li$_2$BeF$_4$ divided by $RT$. HF: approximate conversion with the molar density of the salt. $^b$Extrapolated.\par}
\end{table}

\begin{table}[!htbp]
\caption{Excess chemical potentials (eV), PBE-D3 level.$^a$}
\label{tab:muex}
\begin{ruledtabular}\centering
\begin{tabular}{rcccc}
\hline\hline
$T$ (K) & \Tz & \Tt & TF & He \\
\hline
873  & 0.25 & $0.427 \pm 0.004$ & $-0.05 \pm 0.02$ & 0.482  \\
973  & 0.25 & $0.421 \pm 0.001$ & $-0.01 \pm 0.01$ & 0.482  \\
1073 & 0.26 & $0.416 \pm 0.001$ & $0.00 \pm 0.01$ & 0.480  \\
1173 & 0.27 & $0.407 \pm 0.001$ & $0.00 \pm 0.01$ & 0.476  \\
1273 & 0.27 & $0.399 \pm 0.001$ & $-0.01 \pm 0.01$ & 0.471  \\
\hline\hline
\end{tabular}
\end{ruledtabular}
{\par\vspace{2pt}\footnotesize\raggedright $^a$D3 correction of \Tz\ interpolated at 973 and 1173~K. Errors for He below 0.001~eV.\par}
\end{table}

\subsection*{Henry constants}

\begin{figure}[!htbp]
\centering
\includegraphics[width=0.5\textwidth]{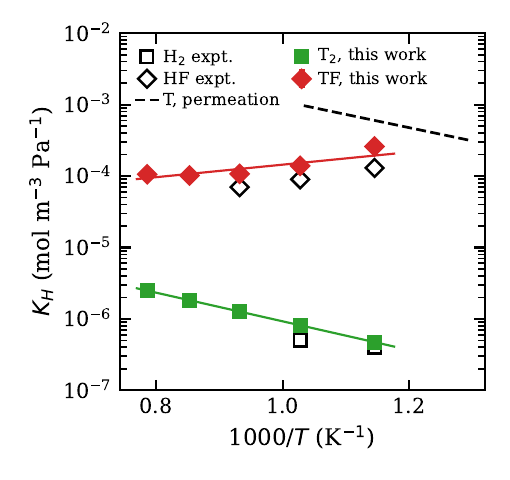}
\caption{Henry constants of \Tt\ and TF from this work, with Arrhenius fits, the measured values for
H$_2$~\cite{Malinauskas1974} and HF~\cite{Field1967}, and the tritium solubility from
permeation~\cite{Calderoni2008}.}
\label{fig:sol}
\end{figure}

The Henry constants of \Tt\ and TF agree with those measured for H$_2$~\cite{Malinauskas1974} and
HF~\cite{Field1967} to within a factor of two (Table~\ref{tab:sol}, Fig.~\ref{fig:sol}), and the
solution enthalpies have the same signs as in experiment. \Tt\ is kept out of the melt by the work
of forming a cavity in a dense ionic liquid, which decreases as the melt expands with temperature.
TF pays a similar cost but recovers it through its charges and its hydrogen bond, and in the
thermodynamic integration the uncharged insertion and the charging step nearly cancel.

\Tt\ does not dissociate. For \Tt\ $\rightleftharpoons$ 2\Tz, the melt changes the gas-phase
equilibrium constant by a factor $\exp[-\beta(2\muex(\Tz) - \muex(\Tt))]$, which with the values of
Table~\ref{tab:muex} lowers it by about a factor of three. Dissociation is already negligible in
the gas at these temperatures, so atomic tritium is a minor species in the salt.

\subsection*{Balance between TF and \Tt}

\begin{figure}[!htbp]
\centering
\includegraphics[width=\textwidth]{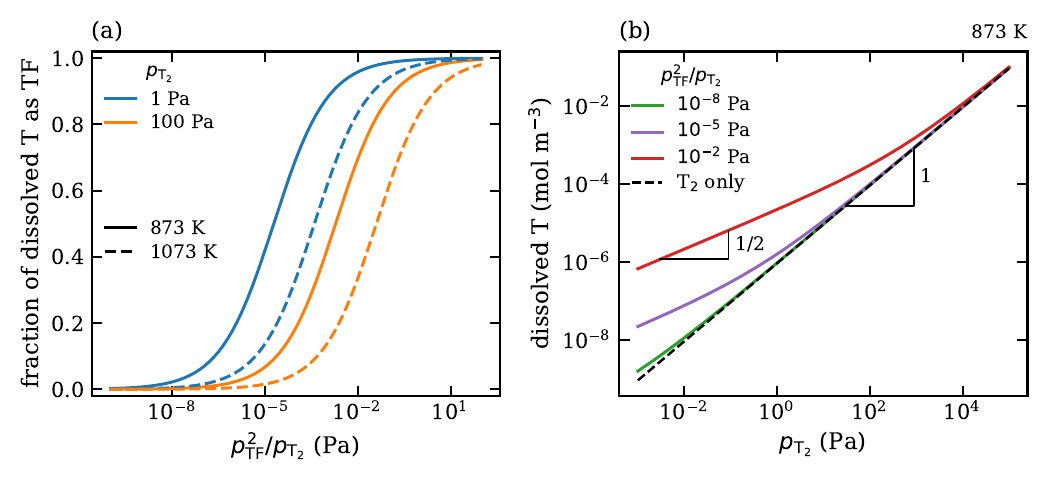}
\caption{Balance between TF and \Tt\ from the Henry constants of this work. (a) Fraction of the
dissolved tritium present as TF against the redox ratio $p_\mathrm{TF}^2/p_\mathrm{T_2}$, for two
T$_2$ pressures at 873 and 1073~K. (b) Dissolved tritium against the T$_2$ pressure at 873~K for
three fixed redox ratios; dashed, \Tt\ alone. The slope is 1 where \Tt\ carries the tritium and 1/2
where TF does.}
\label{fig:redox}
\end{figure}

TF and \Tt\ are linked through the redox state of the salt. If the gas above the salt equilibrates
through T$_2$ + 2F$^-$(salt) $\rightleftharpoons$ 2TF + 2$e^-$(salt), the ratio
\begin{equation}
R = \frac{p_\mathrm{TF}^2}{p_\mathrm{T_2}}
\label{eq:redox}
\end{equation}
is fixed at a given fluorine potential. The amount of tritium cancels in $R$, which therefore
characterizes the redox state alone, large values corresponding to oxidizing and small values to
reducing conditions. With Henry's law for each species, $c_i = K_i p_i$, the fraction of dissolved
tritium present as TF is
\begin{equation}
f_\mathrm{TF} = \left[1 + \frac{2K_\mathrm{T_2}}{K_\mathrm{TF}}\sqrt{\frac{p_\mathrm{T_2}}{R}}\,\right]^{-1}.
\label{eq:fTF}
\end{equation}
Because TF is far more soluble than \Tt, it accounts for half of the dissolved tritium already at
very small $R$ (Fig.~\ref{fig:redox}a), and only a strongly reducing salt, for example one in
contact with metallic Be, keeps the dissolved tritium molecular.

\subsection*{Pressure dependence}

At fixed $R$, $p_\mathrm{TF} = (R\,p_\mathrm{T_2})^{1/2}$, so the dissolved TF increases with the
square root of the \Tt\ pressure while the dissolved \Tt\ increases linearly. The total dissolved
tritium therefore scales as $p_\mathrm{T_2}$ when \Tt\ dominates and as $p_\mathrm{T_2}^{1/2}$ when
TF dominates (Fig.~\ref{fig:redox}b). The local slope in Fig.~\ref{fig:redox}b is $n = 1 -
f_\mathrm{TF}/2$, so a measured pressure exponent between 1/2 and 1 gives the fraction of the
dissolved tritium present as TF. The crossover lies where the two species carry equal amounts, at
$p_\mathrm{T_2} = R\,(K_\mathrm{TF}/2K_\mathrm{T_2})^2$, and moves to higher pressure as the salt
becomes more oxidizing. A salt that is not strongly reduced therefore appears to follow Sieverts'
law, although no molecule dissociates in it.

This is consistent with the permeation experiments of Calderoni \textit{et al.}~\cite{Calderoni2008},
in which T$_2$ in argon entered the salt through a nickel membrane, the downstream side was swept
with argon containing H$_2$, and the redox state was not controlled. The flux varied with the square
root of the T$_2$ pressure, and the derived solubility at 873~K is about twice our TF value and more
than three orders of magnitude above our \Tt\ value (Fig.~\ref{fig:sol}). Dissolved \Tt\ would give
a flux linear in pressure and a much lower solubility, whereas TF formed in the salt accounts for
both the square-root dependence and the magnitude of the solubility. The positive apparent solution
enthalpy then contains the temperature dependence of the reaction that forms TF and is not the
solution enthalpy of TF itself, which we find to be negative.

\subsection*{Helium}

Helium dissolves in the same way as \Tt: it is neutral, does not bond to the melt, and its
solubility is controlled by the cavity term. Malinauskas and Richardson~\cite{Malinauskas1974}
measured He and H$_2$ in the same Li$_2$BeF$_4$ and found He about twice as soluble as H$_2$, and
Watson \textit{et al.}~\cite{Watson1962} reported somewhat higher values in a slightly more
LiF-rich melt. The model reverses this order. He is about half as soluble as \Tt\ at all
temperatures and three to four times less soluble than measured at 873~K (Table~\ref{tab:sol}). Its
solution enthalpy is close to that of \Tt\ and nearly twice the measured value, so the discrepancy
decreases towards 1273~K. The fit reproduces the PBE-D3 energies at the cavity sites closely
(Table~\ref{tab:dfterr}), so the error lies either in the DFT reference or in the model melt. Since
\Tt\ dissolves by the same cavity mechanism, its enthalpy, which is also larger than measured for
H$_2$, may carry the same bias. Direct coexistence with a He bubble, described below, gives a He
solubility about twice that from insertion and halves the discrepancy at 873~K.

\subsection*{Tritium and a helium bubble}

\begin{figure}[!htbp]
\centering
\includegraphics[width=\textwidth]{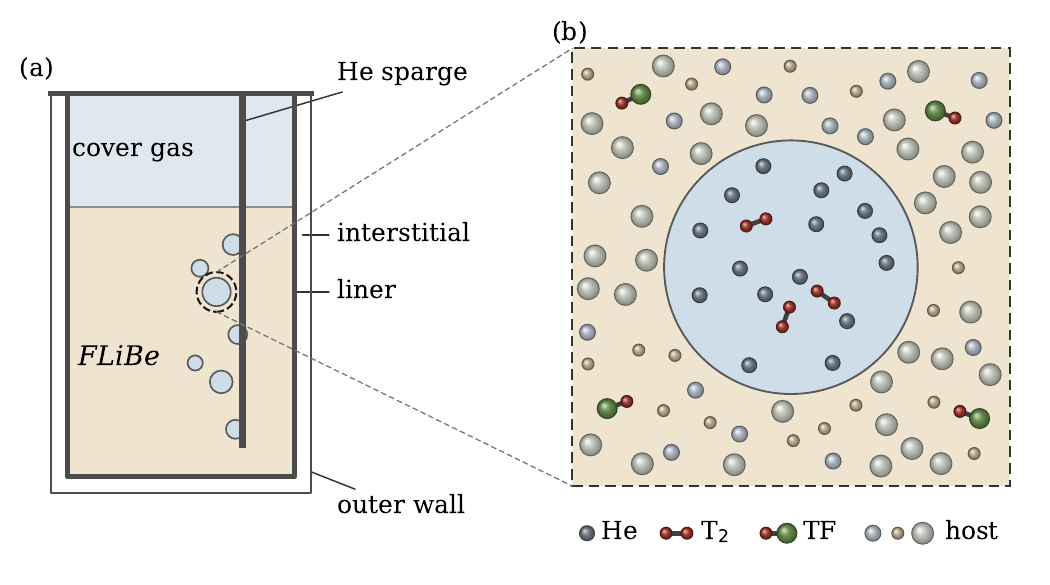}
\caption{(a) A tank with helium sparging, after Ferry \textit{et al.}~\cite{Ferry2023}. (b) The
simulation: a He bubble in periodic FLiBe with \Tt\ and TF.}
\label{fig:bubscheme}
\end{figure}

In LIBRA, tritium is removed mainly by sparging: helium bubbles rise through the salt and are
assumed to leave in equilibrium with it, with Henry's law applied at the bubble
surface~\cite{Ferry2023}. We examined this with a helium bubble embedded in the melt
(Fig.~\ref{fig:bubscheme}). Host ions were removed from a sphere at the centre of a large cell, the
cavity was filled with He, and equal numbers of \Tt\ and TF were placed either in the salt or in
the bubble. Both systems were run at 873~K and 1~bar (Table~\ref{tab:settings}), and the part of each
trajectory after the He distribution had settled was analysed. The bubble is defined as the largest
cluster of He atoms, its radius is obtained from the spread of their positions, and densities are
computed as a function of the distance from its centre.

\begin{figure}[!htbp]
\centering
\includegraphics[width=\textwidth]{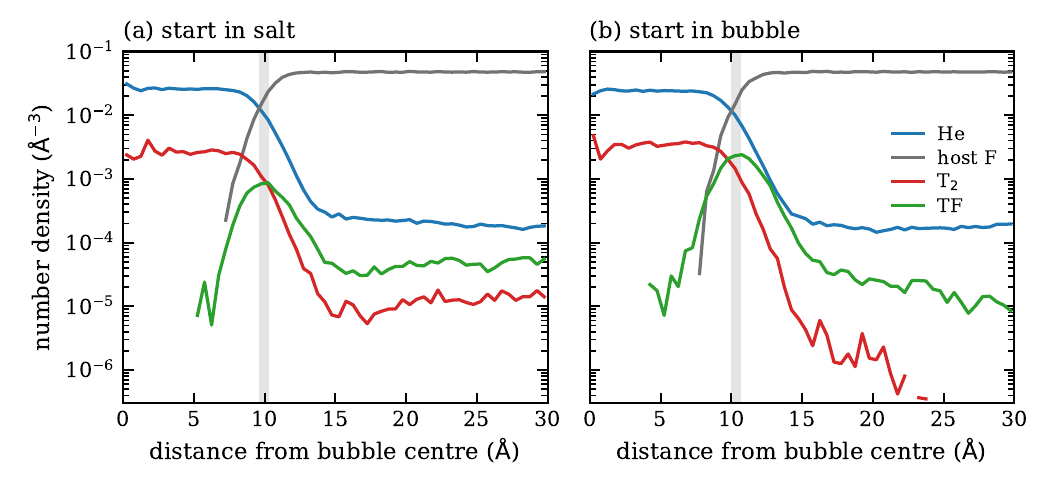}
\caption{Number densities around the He bubble at 873~K, with \Tt\ and TF started (a) in the salt
and (b) in the bubble. The grey band marks the interface, between the point where He falls to half
its value in the bubble and the point where host F reaches half its value in the salt.}
\label{fig:bubprof}
\end{figure}

The bubble remains intact, but about a third of the He dissolves in the salt as isolated atoms
within the first nanosecond, after which the numbers stay constant in both runs. The dissolved He
is distributed uniformly through the salt, with no excess at the interface (Fig.~\ref{fig:bubprof}).
\Tt\ fills the bubble at a nearly uniform density and is depleted in the salt by about two orders of
magnitude. TF is almost absent from the gas, as its Henry constant predicts for a bubble of this
size, which at equilibrium would hold about a tenth of a TF molecule. It accumulates instead at the
salt surface, where its density exceeds the bulk value by one to two orders of magnitude.

The gas in a bubble of nanometre size is at kilobar pressure and far from ideal, so the density
ratio across the interface is not the Ostwald coefficient at 1~atm. The excess chemical potentials
of He and \Tt\ in the bubble were therefore obtained by test insertion into its core and combined
with the density ratio to give the excess chemical potential in the salt (Table~\ref{tab:bubble}).
For \Tt\ the two starting configurations bracket the value obtained by insertion into the pure
melt, so the presence of helium does not change the solubility of \Tt. For He both runs give a solubility about twice that from insertion, halfway to the measured
value. The two routes use the same potential, so the difference must come from the way each samples
the melt. The large cavities that dominate the insertion average for He are one possible cause, but
the origin has not been resolved.

\begin{table}[!htbp]
\caption{Excess chemical potential in the salt and Ostwald coefficient at 873~K, from the bubble
and from insertion into the pure melt.$^a$}
\label{tab:bubble}
\begin{ruledtabular}\centering
\begin{tabular}{llccccc}
\hline\hline
& & \multicolumn{2}{c}{\muex\ (eV)} & \multicolumn{3}{c}{$L$ ($10^{-3}$)} \\
& start & bubble & insertion & bubble & insertion & expt.~\cite{Malinauskas1974} \\
\hline
\Tt & salt   & 0.41 & 0.43 & 4.5 & 3.4 & 3.2$^b$ \\
\Tt & bubble & 0.52 & 0.43 & 1.0 & 3.4 & 3.2$^b$ \\
He  & salt   & 0.44 & 0.48 & 3.1 & 1.7 & 6.0 \\
He  & bubble & 0.43 & 0.48 & 3.5 & 1.7 & 6.0 \\
\hline\hline
\end{tabular}
\end{ruledtabular}
{\par\vspace{2pt}\footnotesize\raggedright $^a$$L$ at 1~atm. $^b$H$_2$.\par}
\end{table}

These results support the treatment of \Tt\ in LIBRA, since a bubble takes up \Tt\ with a partition
close to that given by its Henry constant. They also support the addition of Be to keep the tritium
as \Tt, because TF is far less volatile and is removed much less efficiently by the gas. The Henry's-law description leaves out the excess of TF
at the interface, so a rising bubble also carries some TF on its surface. For bubbles of 0.5~mm this
contribution is small compared with the \Tt\ inside, but the surface is where TF meets the H$_2$ of
the sparge gas, and reactions there lie outside the present model. A movie of the run started with
the tritium in the salt is provided as Supporting Information.

\section{Diffusion}\label{sec:transport}

\begin{figure}[!htbp]
\centering
\includegraphics[width=0.6\textwidth]{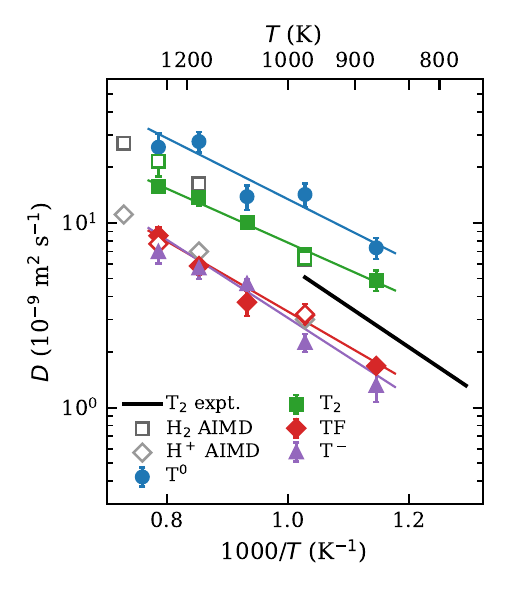}
\caption{Diffusion coefficients of the tritium species against $1/T$, corrected for the cell size
(filled: 108 units; open: 256), with Arrhenius fits. Thick line: tritium in FLiBe from
permeation~\cite{Calderoni2008}. Open grey symbols: H$_2$ and H$^+$ from AIMD~\cite{Lam2021}.}
\label{fig:D}
\end{figure}

The mean squared displacements are linear over the fitting window for all species and
temperatures. The finite-size correction is small for the neutral species and larger for TF and  \Tm{}. The larger cells agree with the corrected small-cell values except for \Tt\ at the highest
temperature, where the large-cell runs scatter widely (Fig.~\ref{fig:D}); with a single tracer the
scatter between runs is larger than the standard error of the small cells suggests.

\begin{table}[!htbp]
\caption{Diffusion coefficients of the tritium species ($10^{-9}$~m$^2$\,s$^{-1}$) and Arrhenius
parameters.$^a$}
\label{tab:D}
\begin{ruledtabular}\centering
\begin{tabular}{rccccc}
\hline\hline
$T$ (K) & \Tz & \Tt & TF & \Tm & expt.$^b$ \\
\hline
873  & $7.3\pm1.0$  & $4.9\pm0.6$  & $1.7\pm0.1$ & $1.3\pm0.3$ & 2.8 \\
973  & $14.3\pm2.1$ & $6.7\pm0.5$  & $3.2\pm0.2$ & $2.3\pm0.2$ & 5.1 \\
1073 & $13.9\pm2.1$ & $10.0\pm0.5$ & $3.7\pm0.6$ & $4.7\pm0.3$ & -- \\
1173 & $27.6\pm3.5$ & $13.7\pm1.3$ & $5.8\pm0.6$ & $5.7\pm0.7$ & -- \\
1273 & $25.7\pm4.9$ & $15.7\pm0.9$ & $8.5\pm0.9$ & $7.0\pm1.0$ & -- \\
\hline
$D_0$ ($10^{-7}$~m$^2$\,s$^{-1}$) & 6.0 & 2.3 & 2.6 & 4.0 & 9.4 \\
$E_D$ (kJ/mol) & $32\pm6$ & $28\pm2$ & $36\pm3$ & $41\pm8$ & 42 \\
\hline\hline
\end{tabular}
\end{ruledtabular}
{\par\vspace{2pt}\footnotesize\raggedright $^a$Corrected for cell size; standard errors over four runs; $D = D_0\exp(-E_D/RT)$, 873--1273~K. $^b$Tritium permeation, Calderoni \textit{et al.}\cite{Calderoni2008}\par}
\end{table}

The results are given in Table~\ref{tab:D} and Fig.~\ref{fig:D}. The diffusivities decrease in the
order \Tz\ $>$ \Tt\ $>$ TF $\approx$ \Tm, with about a factor of five between the fastest and the
slowest. The neutral species move through the free volume of the melt, and \Tz\ diffuses about
twice as fast as the larger and heavier \Tt, although with more scatter between runs. TF is slower
by a further factor of two to three, since it carries its own fluoride and is hydrogen-bonded to a
second one most of the time, but it still moves about twice as fast as the host fluoride. \Tm{} sits in the first shell of Be and diffuses at about the rate of TF. Because the model
viscosity is too low near the liquidus, the tracer diffusivities there are expected to be too high
and their activation energies too low, especially for TF and \Tm{}, which move with the
network.

Table~\ref{tab:D} also lists the permeation diffusivities of Calderoni \textit{et
al.}~\cite{Calderoni2008}, and Fig.~\ref{fig:D} includes the AIMD results of Lam \textit{et
al.}~\cite{Lam2021}. Our \Tt\ diffuses faster than the permeation value, with a lower activation
energy, by about as much as the error in the host viscosity would produce. If \Tt\ followed the
Stokes--Einstein relation, scaling by the ratio of model to measured viscosity would bring both
quantities close to the measurement. A small neutral solute need not obey this relation, however,
and we offer it as an explanation of the difference rather than as a corrected value. TF diffuses
more slowly than the measured value. The AIMD diffusivities of H$_2$ and of fluoride-bound hydrogen
agree closely with our \Tt\ and TF.

The solubility from the permeation experiments matches TF rather than \Tt, whereas the diffusivity
is closer to that of \Tt. A permeation transient gives the diffusivity of the species that carries
the flux. If this species is TF, our TF diffusivity is too low by a factor of about 1.6, and if the
salt interconverts TF and \Tt\ during the transient, the measured value is an average that our
\Tt\ and TF values bracket. The redox state was not controlled in those experiments, and the data
cannot distinguish between these possibilities; a permeation measurement at a defined fluorine
potential would.

\section{Several tritium species in one cell}\label{sec:multi}

The interactions between tritium species were tested at 973~K in large cells containing several
copies of one species (Tables~\ref{tab:settings} and \ref{tab:multi}). Without these interactions
\Tz\ and \Tt\ gathered in the same cavities, and the \Tt\ diffusivity was halved. With them, two
\Tz\ never come within 1~\AA, two \Tt\ rarely meet, and the diffusivities agree with the
single-tracer values within their errors. TF molecules rarely hydrogen-bond to each other, and
\Tm{} ions meet less often than they would if placed at random on fluoride sites, so they avoid each
other slightly.

\begin{table}[!htbp]
\caption{Cells with several copies of one species at 973~K.$^a$}
\label{tab:multi}
\begin{ruledtabular}\centering
\begin{tabular}{lccccc}
\hline\hline
& \multicolumn{2}{c}{4 copies} & \multicolumn{2}{c}{16 copies} & single \\
& no T--T & T--T & no T--T & T--T & tracer \\
\hline
\Tz\ pairs $<$1~\AA\ (\%) & 3 & 0 & 63 & 0 & \\
\Tt\ pairs $<$2~\AA\ (\%) & 49 & 0.1 & 100 & 2.4 & \\
TF H-bonded pairs (\%) & & 0.1 & & 5 & \\
\Tm\ pairs $<$3~\AA\ (\%)$^b$ & & 1.9 & & 27 & \\
$D$(\Tz) & & 11.1 & & 12.4 & 14.2 \\
$D$(\Tt) & & 6.7 & & 6.3 & 6.7 \\
$D$(TF) & & 3.4 & & 2.8 & 3.1 \\
$D$(\Tm) & & 2.5 & & 2.7 & 2.2 \\
\hline\hline
\end{tabular}
\end{ruledtabular}
{\par\vspace{2pt}\footnotesize\raggedright $^a$500 Li$_2$BeF$_4$ units. Pairs: percentage of frames with at least one pair in contact. $D$ in $10^{-9}$~m$^2$\,s$^{-1}$. T--T: interactions between tritium species, Eq.~(\ref{eq:tt}). $^b$About 50\% for 16 \Tm{} ions placed at random on F sites.\par}
\end{table}

\section{Limitations}

The model uses fixed charges and fixed bonds and does not describe redox chemistry. The balance
between TF and \Tt\ and the decomposition of \Tm{} depend on the fluorine potential of the
salt, which has to be specified from outside. Since bonds cannot break, proton transfer between
fluorides, recombination of two \Tz, reaction of \Tm{} with TF and isotope exchange are all
absent. The DFT scans of hydrogen between two fluorides support the neglect of the first, but the
others would require a reactive model. Nuclei are treated classically, so isotopes differ only in
mass; path-integral simulations with the same potential would give the quantum corrections, which
are largest for H and smallest for T. Gas--salt and metal--salt interfaces and impurities were not
part of the fit, and the He--tritium functions are based on gas-phase pairs only. The rigid-ion host
underestimates the thermal expansion and, with its network corrected to DFT, is too fluid near the
liquidus; the cation--cation correction was also made at a single temperature. A polarizable host
would address the viscosity.

\section{Conclusions}

We have developed a fixed-charge classical potential, fitted only to DFT, for molten FLiBe
containing atomic tritium, T$_2$, TF, \Tm{} and helium. The calculated solubilities of T$_2$
and TF agree with experiment, and neither molecule dissociates in the melt. TF is nevertheless so
much more soluble that it carries most of the dissolved tritium unless the salt is strongly
reducing, and at a fixed redox state its concentration follows the square root of the T$_2$
pressure. This accounts for both the pressure dependence and the magnitude of the solubility
obtained from permeation experiments.

Helium does not change the solubility of T$_2$. In simulations of a helium bubble in the melt,
T$_2$ divides between gas and salt as its Henry constant predicts, while TF, which is far less
volatile, accumulates at the interface. Recovery of tritium by helium sparging therefore depends on
keeping the tritium as T$_2$, that is, on control of the redox state of the salt. Little helium
dissolves at sparging pressures, and the model underestimates its solubility.

The diffusivities are harder to reconcile with experiment. Our T$_2$ diffuses faster than the
permeation value, partly because the model viscosity is too low near the liquidus, while TF and
\Tm{} diffuse two to three times more slowly than T$_2$. Whether the permeation flux is
carried by TF, by T$_2$, or by both through interconversion can only be decided by measurements at
a defined redox state.

With its cation--cation interactions corrected against neural-network MD, the rigid-ion host
reproduces the density and the measured neutron structure factor, and DFT supports a fixed T--F
bond and places \Tm{} next to Be. The potential is released as LAMMPS input files for any
combination of these species.

\appendix
\section{Appendix: Energy expressions}\label{app:energy}

The energy of each species is given below. $Y_i$ is the species (Li, Be or F) of host atom $i$,
$q_i$ its charge, and $V_{\mathrm{F}Y}$ the host F--$Y$ function of Eq.~(\ref{eq:bmh}). $U_{XY}$
are functions of the form of Eq.~(\ref{eq:rbf}), each continued below its shortest training
distance as described in the section \textit{Tritium--host interactions}. The Coulomb sums are
evaluated with PPPM.

\paragraph{Host.}
\begin{equation}
U_\mathrm{host} = \sum_{i<j}\left[\frac{q_iq_j}{4\pi\varepsilon_0 r_{ij}} + V_{ab}(r_{ij}) + \Delta V_{ab}(r_{ij})\right],
\end{equation}
with $q_\mathrm{Li} = 0.908$, $q_\mathrm{Be} = 1.816$ and $q_\mathrm{F} = -0.908$\,e. $V_{ab}$ is
Eq.~(\ref{eq:bmh}) with the parameters of Table~\ref{tab:host}, and $\Delta V_{ab}$ is the
tabulated correction of Eq.~(\ref{eq:ibi}) for Li--Li, Be--Li and Be--Be and zero for the other
pairs.

\paragraph{\Tz.} A neutral atom at $\mathbf{r}_\mathrm{T}$,
\begin{equation}
U_{\mathrm{T}^0} = \sum_{i} U_{\mathrm{T}^0 Y_i}(r_{\mathrm{T}i}).
\end{equation}

\paragraph{\Tt.} Two neutral atoms T$_1$ and T$_2$ joined by a Morse bond,
\begin{equation}
U_{\mathrm{T}_2} = V_\mathrm{M}(r_{12}) + \sum_{a=1,2}\sum_{i} U_{\mathrm{T}Y_i}(r_{ai}),
\qquad V_\mathrm{M}(r) = D\left[1 - e^{-a(r - r_e)}\right]^2,
\end{equation}
with $D = 4.57$~eV, $a = 1.99$~\AA$^{-1}$ and $r_e = 0.752$~\AA.

\paragraph{TF.} The tritium carries $+q$ and its fluoride \Fb\ carries $-q$, with $q = 0.675$\,e,
\begin{equation}
U_\mathrm{TF} = V_\mathrm{M}(r_{\mathrm{TF_b}})
 + \sum_{i}\left[\frac{q\,q_i}{4\pi\varepsilon_0 r_{\mathrm{T}i}} + U_{\mathrm{T}Y_i}(r_{\mathrm{T}i})\right]
 + \sum_{i}\left[\frac{-q\,q_i}{4\pi\varepsilon_0 r_{\mathrm{F_b}i}} + V_{\mathrm{F}Y_i}(r_{\mathrm{F_b}i})
   + U_{\mathrm{F_b}Y_i}(r_{\mathrm{F_b}i})\right],
\end{equation}
with $D = 5.98$~eV, $a = 2.13$~\AA$^{-1}$ and $r_e = 0.935$~\AA. The Coulomb and short-range
interactions between the two atoms of the molecule are excluded, and only the Morse bond acts
between them.

\paragraph{\Tm.} \Tm{} replaces a fluoride and carries its charge, $q_{\mathrm{T}^-} = -0.908$\,e,
\begin{equation}
U_{\mathrm{T}^-} = \sum_{i}\left[\frac{q_{\mathrm{T}^-}q_i}{4\pi\varepsilon_0 r_{\mathrm{T}i}}
 + U_{\mathrm{T}^-Y_i}(r_{\mathrm{T}i})\right].
\end{equation}

\paragraph{Helium.} A neutral atom with functions to the host, the HFD-B function to other He,
and site--site functions to the tritium species,
\begin{equation}
U_\mathrm{He} = \sum_{i} U_{\mathrm{He}Y_i}(r_{\mathrm{He}i}) + \sum_{\mathrm{He}'} V_\mathrm{HFD\text{-}B}(r_{\mathrm{HeHe}'})
 + \sum_{s} u_{\mathrm{He}s}(r_{\mathrm{He}s}),
\end{equation}
where $s$ runs over \Tz, the atoms of \Tt\ and TF, and \Tm{}, and $u_{\mathrm{He}s}$ has the
form of Eq.~(\ref{eq:tt}) except for \Tm{}, which uses $U_{\mathrm{He}\,\mathrm{F}}$.

\paragraph{Between tritium species.} For two sites $s$ and $t$ on different species,
\begin{equation}
U_{\alpha\beta} = \sum_{s\in\alpha}\sum_{t\in\beta}\left[\frac{q_sq_t}{4\pi\varepsilon_0 r_{st}} + u_{st}(r_{st})\right],
\end{equation}
where $u_{st}$ has the form of Eq.~(\ref{eq:tt}) for pairs involving \Tz\ or \Tt, and is the host
function of the closest analogue for pairs between TF molecules and pairs involving \Tm{}.

\begin{suppinfo}
DFT data sets; host fitting sequence; all six partial distribution functions; TF charge scan and
short-range continuations; stability tests and sampling of TF; Be--F--T relaxations; gas-phase
interactions between tritium species; D3 corrections; thermodynamic integration; Wannier analysis
of \Tm{}; helium fit and comparison of the He--\Tz\ interaction with earlier work; use of the
potential; movie of the helium bubble simulation (MP4).
\end{suppinfo}

\begin{acknowledgement}
This work is part of the UKAEA’s Lithium Breeding Tritium Innovation (LIBRTI) programme, funded by the Department of Energy Security and Net Zero, as announced by the UK Government in October 2023. DNM work has been funded by the Engineering and Physical Sciences Research Council (EPSRC) Fusion Grant 2022/27 [grant number EP/W006839/1]. DNM, and DA acknowledge EUROfusion for access to the PITAGORA high-performance computing facilities at CINECA, Bologna, Italy. 
\end{acknowledgement}

\section*{Conflict of interest}
The authors declare no competing financial interest.

\section*{Data and Software Availability}
The potential and examples will be published on Zenodo upon publication of this work.

\bibliography{refs}

\end{document}